\documentclass[12pt]{article}
\usepackage{algorithm,rotating,verbatim,tabularx}
\usepackage[authoryear,round]{natbib}
\RequirePackage{hyperref}
\usepackage{amsmath,latexsym,amssymb,graphicx,amsthm}

\usepackage{setspace}
\DeclareMathOperator{\Ps}{\mathbf{\Psi}}
\DeclareMathOperator{\Thet}{\mathbf{\Theta}}

\DeclareMathOperator{\A}{\mathbf{A}}

\DeclareMathOperator{\Y}{\mathbf{Y}}

\DeclareMathOperator{\X}{\mathbf{X}}

\DeclareMathOperator*{\minimize}{\mathrm{minimize}}
\DeclareMathOperator*{\argmin}{\mathrm{argmin}}
\DeclareMathOperator*{\maximize}{\mathrm{maximize}}

\DeclareMathOperator{\vrho}{\boldsymbol{\rho}}

\begin{document}

\title{A Log-Linear Graphical Model for
  Inferring Genetic Networks from High-Throughput Sequencing Data}

\author{Genevera I. Allen$^{1,2}$ \& Zhandong Liu$^{1}$ \\ 
{\small $^{1}$ Department of Pediatrics-Neurology, Baylor
  College of Medicine} \\
{\small \& Jan and Dan Duncan Neurological Research Institute at
  Texas Children's Hospital,} \\ {\small $^{2}$ Department of
  Statistics, Rice 
  University. }}

\maketitle

\begin{abstract}
Gaussian graphical models
are often used to infer gene networks based on microarray expression
data.  Many scientists, however, have begun using high-throughput sequencing
technologies to measure gene expression.  As the resulting
high-dimensional count data consists of counts of
sequencing reads for each gene, Gaussian graphical models are not
optimal for modeling gene networks based on this discrete data.
 We develop a novel method for estimating high-dimensional
Poisson graphical models, the 
{\em Log-Linear Graphical Model}, allowing us to infer networks based on
high-throughput sequencing data.  Our model assumes a pair-wise Markov
property: conditional on all other variables, each variable is
Poisson.  We estimate our model locally via neighborhood selection 
by fitting $\ell_{1}$-norm penalized log-linear models.  Additionally,
we develop a fast parallel algorithm, an approach we
call the {\em Poisson Graphical Lasso}, permitting us to fit our
graphical model to high-dimensional genomic data sets.
In simulations, we illustrate the effectiveness of our methods for
recovering network structure from count data.  A case study on breast
cancer microRNAs, a novel application of graphical models, finds 
known regulators of breast cancer genes and discovers novel microRNA
clusters and hubs that are targets for future research. \\
{\bf Keywords}: Markov networks, graphical models, Hammersley-Clifford
theorem, next generation sequencing data, 
microRNAs, regulatory networks
\end{abstract}

\section{Introduction}

Graphical models have become a popular technique to depict and explore
relationships between genes and estimate genomic pathways
\citep{dobra_2004, krmer_2009}.
Undirected 
graphical models, or Markov Networks, denote conditional dependence
relationships between genes \citep{dempster_1972}.   In other words,
genes A and B are linked if given the profiles across the subjects for
all other genes, the levels of gene A 
are still predictive of the levels of gene B.  Thus, Markov Networks
denote a type of direct dependence that is stronger than merely
correlated expression values.  Many have developed methods
to estimate high-dimensional Markov Networks for Gaussian or binary
data by using sparsity to select the edges between genes
\citep{meinshausen_graphs_2006, 
  yuan_lin_2007, glasso,  ravikumar_ising_2010}. 
Several have used these methods for Gaussian graphical models to infer
network structures from microarray gene expression 
data \citep{meinshausen_graphs_2006, liu_STARS_2010}.  As
typical log ratio expression values from microarray 
data follow approximately a Gaussian distribution, these models are
appropriate.  Recently, more scientists are using RNA-sequencing
technologies to measure gene expression or microRNA levels, as these
methods in theory 
yield less technological variation than that of microarrays
\citep{marioni_2008}. 
Measurements from RNA-sequencing, however, are not approximately
Gaussian and are in fact read counts of how many times a transcript
has been mapped to a specific genomic location.  The RNA 
sequencing expression values are then integer valued and non-negative;
thus, many have 
advocated to model this count data using the Poisson distribution
\citep{marioni_2008, bullard_rna_2010, li_rna_2011}.
In this paper, we develop a novel Log-Linear Graphical Model 
based on the Poisson distribution and build an algorithm to estimate
genomic networks from high-dimensional sequencing data.


High-dimensional methods to estimate Gaussian or binary Markov
Networks have been well-studied \citep{meinshausen_graphs_2006,
  yuan_lin_2007, glasso,
  ravikumar_ising_2010}.  Few, however, have 
introduced methods for Poisson 
Markov Networks.  \citet{esl_2nd} propose a combinatorial approach, 
augmenting the data matrix and fitting log-linear regression models.
The computational complexity of the method, however, grows on the
order of $p^{2}2^{p}$ where $p$ is the number of variables; this is
infeasible for data with $p>20$. Others have outlined related
approaches for multi-way contingency tables
\citep{madigan_1995, lauritzen_1996, bishop_discrete_2007,
  wainwright_jordan_2008} that again, are  
only computationally feasible for a small number of
variables.  Our goal in this
paper is to develop a model and computationally attractive algorithm
to estimate high-dimensional 
Poisson graphical models that can be used to infer genetic networks
based on RNA-sequencing data.

In this paper, we make the several novel contributions: (1)
propose a Log-Linear Graphical Model based on a pair-wise Poisson
Markov Network; (2) introduce a neighborhood selection approach to
infer network structure locally  via a series of $\ell_{1}$ penalized
log-linear 
models; and (3) build a fast parallel algorithm to fit 
our graphical model to high-dimensional genomic data.  These methods
are developed in detail in Section \ref{section_methods}.  In Section
\ref{section_sims}, we study the 
utility of our methods for recovering the underlying graph structure
of simulated Poisson networks.  We apply our LLGM to infer
high-dimensional networks for breast cancer microRNAs, a novel applied
contribution, in Section \ref{section_miRNA}. In Section
\ref{section_dis}, we conclude with a 
discussion of 
the implications of our work and directions for future research.

\section{Methods}
\label{section_methods}

We develop a log-linear graphical model and a fast algorithm to fit
this model, the Poisson Graphical Lasso, that can be used to estimate
genomic networks 
from high-dimensional sequencing data.  We begin with background and
considerations for modeling sequencing data using the Poisson
distribution.

\subsection{Poisson Models for Sequencing Data}
\label{section_Pois}

High-throughput RNA-sequencing technologies quantify expression values
by mapping short reads of cDNA back to the original genome.  The
resulting data consist of counts at each genomic location that are
non-negative integers \citep{marioni_2008}.  Several
models have 
been proposed for 
this data including Poisson and negative binomial models
\citep{anders_2010, robinson_2010, li_rna_2011}.  In 
this paper, we will assume that after normalization the data follows a
Poisson distribution with a separate mean for each gene.

There are several items one must consider when normalizing
high-throughput sequencing data.  First, the samples may contain
vastly different numbers of total read counts reflecting technological
variation in sequencing depths with no biological relevance
\citep{marioni_2008, mortazavi_2008}.
Some have suggested to normalize samples by the total counts
\citep{marioni_2008}, 
the RPKM (reads per KB per million) \citep{mortazavi_2008,
  jiang_rna_2009}, or more robust 
methods such as 
normalizing via the geometric mean \citep{anders_2010} or quantiles
\citep{bullard_rna_2010}. 
Another characteristic of sequencing data is that the read counts for
some genomic locations may have zero or nearly zero expression
values \citep{mortazavi_2008}.  As these genes and others that are
constant across 
the samples will not be meaningful genes to study via network models,
we filter out these genes.  Finally, many have noted that even after
adjusting for sequencing depth and filtering genes, the data can still
be overdispersed compared to a Poisson distribution \citep{robinson_2010,
  anders_2010, li_rna_2011}.  We 
choose to adjust for this by transforming the data via a power $\alpha
\in (0,1]$ where 
  $\alpha$ is chosen to yield approximately Poisson data
  \citep{li_rna_2011}.  While others have advocated using the negative
  binomial distribution in this context \citep{robinson_2010,
  anders_2010}, the Poisson
  distribution has a number of advantages for graphical models.  These
  include a simple one-parameter form and well established methods for
  fitting penalized log-linear models.

In summary we employ three common steps to normalizing high-throughput 
  sequencing data: (i) adjust for sequencing depth, (ii) filter out
  genes with low variance across samples, and (iii) adjust for
  possible overdispersion. 
After this normalization, we assume that the data follows a Poisson
distribution with a separate mean for each gene.  Specifically, if we have
sequencing data $X_{ij}$ for $i = 1, \ldots n$ samples and $j = 1,
\ldots p$ genes, then we assume that for each gene $j$: $X_{1,j},
  \ldots X_{n,j} \ \  \overset{iid}{\sim} \ \  Poisson( \lambda_{j}
  )$. 


\subsection{A Log-Linear Graphical Model}

We develop the mathematical framework for our Log-Linear Graphical
Model by assuming pair-wise conditional Poisson relationships between
variables and defining a Poisson Markov Network.  Interestingly, this
joint distribution on the nodes places severe constraints on the types
of pair-wise conditional relationships between variables that have
little meaning in the context of genetic networks.  Thus, we propose
to estimate 
unrestricted relationships between pairs of variables locally
and then put together this set of local networks.  While this
procedure does not fit a joint model on all the nodes, it will allow
us to infer a more
general graph structure.   To denote the difference between this
set of local models and the joint Poisson Markov Network, we use {\em
  llgm} and {\em LLGM} respectively.

First, let us define the
undirected network structure as $\mathcal{G} = \{ V, E \}$;
that is, the graph, $\mathcal{G}$, consists of the set of vertices
(variables), $V$, and the set of edges (links), $E$.  In the context
of genetic networks, each of the vertices or nodes corresponds to a specific
gene and edges denote links or important relationships between genes.  
Let $\X$ be a matrix of $n$ samples measured on $p$ genes with
entries taking on integer values, $X_{ij} \in \{0, 1, 2, \ldots
\infty \}$.

Our model is characterized by conditional Poisson
relationships between pairs of nodes: $X_{j}
| X_{k} \  \sim \ Poisson \ \ \forall \ j \neq k \in V$,
where $X_{j}$ is the vector of $n$ samples measured for variable $j$.
Then, both the {\em llgm} and {\em LLGM} models are defined in terms of this
conditional density for each node:
{\small
\begin{align}
\label{llgm}
p( X_{j} | X_{k} = x_{k}  \ \forall k \neq j , \Thet ) \  \sim \
Poisson \left( e^{ \theta_{j} + \sum_{k \neq j}
\theta_{jk} x_{k}} \right),
\end{align}}
with parameters $\Thet = ( \theta_{j}, \theta_{jk}, \ \forall \ j
\neq k \in V )$ where $\theta_{j}, \theta_{jk} \in \Re$.  Here, the
parameter $\theta_{j}$ is an 
intercept, adjusting the conditional mean of $X_{j}$, and the parameter
$\theta_{jk}$ gives the conditional relationship between nodes $j$ and
$k$.  Note that \eqref{llgm} denotes pair-wise relationships or the
local Markov property \citep{lauritzen_1996}.

Recall that for the local, pair-wise Markov property to hold jointly for
all nodes to form a Poisson Markov Random field, the conditions of the
Hammersley-Clifford theorem must be satisfied \citep{hammersley_1971}.
This result 
states that the probability density of the graph may be factored into a
product of potentials on the set of maximal cliques.  \citet{besag_1974}
went on to define the conditions under which Markov random fields can
be defined for exponential families.  From this, the probability
density of the 
Poisson Markov Network, we call this our global {\em LLGM}, is
given by the following:
\begin{align}
\label{LLGM}
p( \X | \Thet  ) = \mathrm{exp} \Big[  \sum_{j \in V} \left( \theta_{j}
  X_{j}  - \mathrm{log}(   X_{j} ! ) \right) 
 + \sum_{(j,k) \in E} 
  \theta_{jk} X_{j} X_{k}   - \Ps( \Thet )  \Big].
\end{align}
Here, $\Ps( \Thet) $ is the log-partition
function: $\Ps( \Thet ) = \mathrm{log} \bigg[  \sum_{x_{j} \in \{0,1,
    \ldots \infty\}, x_{k} \in \{0,1, \ldots \infty\} } \mathrm{exp}
  \big( \sum_{j \in V} ( \theta_{j} x_{j} - \mathrm{log}( x_{j}!) ) +
  \sum_{j,k \in E} \theta_{jk} x_{j} x_{k} \big) \bigg]$. 
This term acts as a normalizing constant ensuring that \eqref{LLGM} is
a proper
probability density function; that is, it sums to one.  This requires
that $\Ps( 
\Thet) < \infty \ \forall \ \X$.  As the term $\theta_{jk} X_{j} X_{k}$
dominates the above summation, this implies that $\theta_{jk} \leq 0 \
\forall \ j \neq k$ \citep{besag_1974}.   Thus, the Poisson Markov
Network, {\em LLGM}, is only defined for conditionally
negatively correlated relationships between variables.

Due to this restriction on the parameter space of the {\em LLGM}, this
model has limited applicability.  Consider for example, that graphical
models are often used to estimate regulatory pathways from gene
expression data.  Thus, conditional on other genes, genes belonging to
the same regulatory pathway would have a positively correlated
relationship.  These positive dependencies cannot be captured or
estimated via the {\em LLGM} model.  Hence, this model is not ideal
for estimating network structures from high-throughput genomic data.

Therefore, we propose to estimate the local model, {\em llgm}, for
each node and then combine each of these estimated local models to
infer a network structure.  Returning to \eqref{llgm},  we will assume
that the intercept term, 
$\theta_{j}$, is zero as we assume each genomic variable has been
adjusted for sequencing depth  as
described in the previous section.  We are then left with the
following log-linear model, the inspiration for the name {\em llgm}: 
\begin{align}
\mathrm{log} \left[  \mathrm{E} ( X_{j} | X_{k} = x_{k}  \ \forall k
  \neq j ) \right] =\sum_{k \neq j} \theta_{jk} x_{k}.
\end{align}
Estimating the parameters of this model, $\theta_{jk} \ \forall \ (j,k)
\in V$, is sufficient for inferring the local network structure of the
graph.  Notice, for example, that if $\theta_{jk} = 0$, then this
implies that $X_{j} \perp X_{k} | X_{\neq j,k}$.  In other words,
variables $j$ and $k$ are conditionally independent given all other
variables.  In our graph structure, $\mathcal{G}$, this local conditional
independence implies that there is no edge between $X_{j}$ and
$X_{k}$.  This approach is closely related to
the neighborhood selection problem proposed for Gaussian graphical
models and Ising models in \citet{meinshausen_graphs_2006} and
\citet{ravikumar_ising_2010} 
respectively.  The major difference between our approach for Poisson
graphical models and these existing methods is that estimating the
pair-wise conditional dependencies for the Ising and Gaussian
graphical model are sufficient for determining the joint dependence
structure of the random field.  While this does not hold for our
models, our local approximation will allow us to estimate a richer set
of dependence structures.

\subsection{Poisson Graphical Lasso}

We describe our method for fitting the local Log-Linear
Graphical Model, {\em llgm}, an algorithm we call the {\em Poisson
  Graphical 
  Lasso} named after the Graphical Lasso algorithm of \citet{glasso} for
Gaussian Graphical Models.

\subsubsection{Neighborhood Selection}

We propose to fit the local Log-Linear Graphical Model, {\em llgm}, by
for each node, 
estimating the set of edges extending out from the node, of the node's
{\em neighborhood}.  
\citet{meinshausen_graphs_2006} first proposed to
automatically select the neighborhood of node $j$ by placing an
$\ell_{1}$-norm 
penalty on linear regression coefficients to encourage sparsity.
The regression coefficients of variables with weak relationships
to variables $j$ will be shrunk to zero, and  there is no edge between
the nodes in the graph.  Variables with strong relationships with
gene $j$ will have non-zero regression coefficients, and these
will be connected to node $j$ in the graph. 
Neighborhood selection methods have been 
developed for high-dimensional graph estimation using $\ell_{1}$-norm
penalized linear regression \citep{meinshausen_graphs_2006} and
logistic regression \citep{ravikumar_ising_2010}.  We extend this to
$\ell_{1}$-norm penalized log-linear regression for neighborhood
selection for the local {\em llgm}.

Mathematically, we can write the neighborhood selection problem for
node $j$ as the solution to the following penalized log-linear
regression problem:
\begin{align}
\label{opt_prob}
\maximize_{\Theta_{\neq j,j}} \   \ \frac{1}{n} \sum_{i=1}^{n} \left[
    X_{ij} \left( X_{i, \neq j} \Theta_{\neq j, j} \right) -
    \mathrm{exp} \left( X_{i, \neq j} \Theta_{\neq j, j}  \right)
    \right]  \  - \rho || \Theta_{\neq j,j} ||_{1}.
\end{align}
Here, $\rho \geq 0$ is a regularization parameter controlling the
amount of sparsity in the neighborhood and the notation $X_{i,\neq
  j}$ denotes the $\i^{th}$ row of $\X$ and all columns other than
column $j$, and analogously for $\Theta_{\neq j,j}$.  Thus, we
estimate the zero elements 
in one column of our parameter matrix, $\Thet$, at a time by
regressing the $j^{th}$ variables, $X_{j}$ onto all other variables $X_{\neq
  j}$.  For ease of notation, we denote this estimated column as
$\hat{\Thet}_{j}(\rho)$ to make explicit the dependency on the
regularization parameter, $\rho$; note that there is no $j^{th}$
element to this column vector.  We denote the estimated graph
structure as the adjacency matrix $\hat{\A}( \rho)$ implied by the
zero elements in $\hat{\Theta}(\rho): \hat{\A}(\rho) = | \mathrm{sign}
( \hat{\Thet}(\rho)) |$.
There are many fast computational
approaches to fitting these $\ell_{1}$-norm penalized log-linear
models \citep{glmnet_paper_2010} that we will
discuss further when we introduce our 
algorithm subsequently.

Finally, notice that neighborhood selection is not symmetric.  In
other words, while nodes $j$ and $k$ may be estimated to have an edge when
node $j$ is regressed on all others, this edge may not be present when
node $k$ is the regressor \citep{meinshausen_graphs_2006,
  ravikumar_ising_2010}.  Thus, we define our estimated 
graph, $\hat{\A}(\rho)$, as the union over the set of these edges,
noting that the intersection is also appropriate:
{\small 
\begin{align}
\label{adj}
\hat{A}_{jk}(\rho) = \mathrm{max} \bigg\{ | \mathrm{sign}( \hat{\Theta}(\rho)_{jk} ) |,
| \mathrm{sign}( \hat{\Theta}(\rho)_{kj} ) |  \bigg\} \ \ \forall \ j
\neq k. 
\end{align}}
In other words, an edge connecting nodes $j$ and $k$ is estimated if either
solving \eqref{opt_prob} with $X_{j}$ or $X_{k}$ as regressors yields
a non-zero coefficient in the other.

\subsubsection{Selecting Regularization Parameters}

The regularization parameter $\rho$ controls the sparsity of the graph
structure, or in other words, the number of estimated links between
nodes.  We seek a data-driven method for estimating this parameter.
In the Gaussian graphical 
model literature, many data-driven methods such as cross-validation,
BIC, AIC, and stability selection have
been proposed.  The former three
approaches, however, require calculating the log-likelihood; recall
that our local algorithm based on the {\em llgm} does not maximize the
joint likelihood of \eqref{LLGM}.   We then, propose to estimate the
regularization parameter 
via stability selection, an approach which seeks the $\rho$ leading
to the most stable set of edges \citep{liu_STARS_2010}.  In brief, stability 
selection sub-samples the data $\X^{(b)}$ and estimates a separate
graph $\A^{(b)}(\vrho)$ for each sub-sample and vector of
regularization parameters, $\vrho$.   The optimal value of $\rho$
controls the average variance over the edges of the sub-sampled graphs
\citep{liu_STARS_2010} (reproduced using our notation for completeness):
{\small 
\begin{align}
\label{stab_select}
\rho_{opt} = \argmin_{\rho} \Bigg\{ \minimize_{0 \leq
  \lambda \leq \rho} \bigg\{ \sum_{j < k} 2
\bar{A}_{jk}(\lambda) ( 1 - \bar{A}_{jk}(\lambda) ) / \left(
\begin{array}{c} p \\ 2 \end{array} \right) \bigg\} 
  \leq \beta \Bigg\},
\end{align}}
where $\bar{A}_{jk}(\vrho) = \frac{1}{B} \sum_{b=1}^{B}
A_{jk}^{(b)}(\vrho)$. We note that default values for $\beta$, $\beta
= .05$,
and the number of sub-samples, $m = \lfloor 10 \sqrt{n} \rfloor$, from
\citep{liu_STARS_2010} are used.

\subsubsection{Algorithm}



We are interested in developing a fast algorithm to fit our {\em llgm}
model to
high-throughput genomic data.  We accomplish this by incorporating
fast path-wise algorithms and stability selection into a
parallel computing framework.

First, notice that each of the penalized log-linear models,
\eqref{opt_prob}, can be fit independently as the results of each do
not depend on others.  Thus, the neighborhood of each node can be
estimated in parallel.  In addition, recent advances in computing
$\ell_{1}$ penalized models via path-wise coordinate methods over a
range of regularization parameters, $\vrho$, 
allow us to compute the entire neighborhood solution path for each
node with approximately the same speed as fitting at a single value of
$\rho$ \citep{glmnet_paper_2010}.  Thus, we seek to fit the penalized
log-linear 
models path-wise over a range of regularization parameters in parallel
for each node.  To accomplish this, the vector of regularization
parameters $\vrho$ we consider must be fixed in advance for each
node.  This means we must know the value of $\rho_{max}$ at which all
coefficients are zero for all nodes, or in other words, no
edges are estimated in the graph.  Examining the Karush-Kuhn-Tucker
conditions of \eqref{opt_prob}, the minimum value of $\rho$ at which
no edges are selected for $X_{j}$ is $\mathrm{max}_{k \neq j} |
X_{k}^{T} X_{j} |$.  Hence, $\rho_{max} =
\mathrm{max}_{j, k \neq j} | X_{k}^{T} X_{j} |$, the maximum over all
the $j$ regression problems.

\begin{algorithm}
\caption{Poisson Graphical Lasso for {\em llgm}}
\label{alg_pglasso}
\begin{enumerate}
\item Normalize the data $\X$ as described in Section 2.1.  Set
  $\rho_{max} = 
  \mathrm{max}_{k,j} | X_{k, \neq j}^{T} X_{j} |$. Fix
  $\rho_{min} \approx 1.0 \times 10^{-4}$. Define 100 log-spaced
  values $\vrho = [\rho_{max} \ \ldots \ \rho_{min} ]^{T}$.
\item For each $X_{j}$, $j = 1, \ldots p$, do in parallel:
\begin{enumerate}
\item Solve \eqref{opt_prob} with regressor $X_{j}$ and predictors
  $X_{\neq j}$ path-wise for $\vrho$ yielding
  $\hat{\Theta}_{j}(\vrho)$. 
\item For $b = 1, \ldots B$:
\begin{enumerate}
\item[i.] Sample $m = \lfloor 10 \sqrt{n} \rfloor$ observations, yielding the
  sub-sampled data, $\X^{(b)}$.
\item[ii.] Solve \eqref{opt_prob} with regressor $X^{(b)}_{j}$ and predictors
  $X^{(b)}_{\neq j}$ path-wise for $\vrho$ yielding
  $\hat{\Theta}^{(b)}_{j}(\vrho)$.  
\end{enumerate}
\end{enumerate}
\item Determine the graphs $\hat{\A} (\vrho)$ from $\hat{\Thet}(\vrho)$
  and $\hat{\A}^{(b)}(\vrho)$ from $\hat{\Thet}^{(b)}(\vrho)$ via \eqref{adj}.
\item Determine $\rho_{opt}$ via stability selection,
  \eqref{stab_select}.
\item Return the graph, $\hat{\A}(\rho_{opt})$.
\end{enumerate}
\end{algorithm}

Algorithm \ref{alg_pglasso} summarizes these items and the steps of
our Poisson Graphical Lasso method.  Notice that the entire set of
computations including path-wise log-linear models and stability
selection are performed in parallel for each node.  This dramatically
reduces the computational complexity to approximately $O((1+B)p^{3})$
for each node \citep{glmnet_paper_2010}.  After this, stability
selection results 
are combined to estimate the optimal regularization parameter and the
final graph is determined via maximum edge agreement.  Thus, our
Poisson Graphical Lasso Algorithm is a computationally efficient
method for inferring the high-dimensional {\em llgm}.

\section{Results}

We evaluate the performance of our local Log-Linear Graphical Model for
recovering network structure via experiments on simulated data and
through a novel 
application to microRNA sequencing data.

\subsection{Experiments on Simulated Networks}
\label{section_sims}

\begin{figure*}[!t]
\includegraphics[width=7.25in,clip=true,trim=.1in 0in 0in 0in]{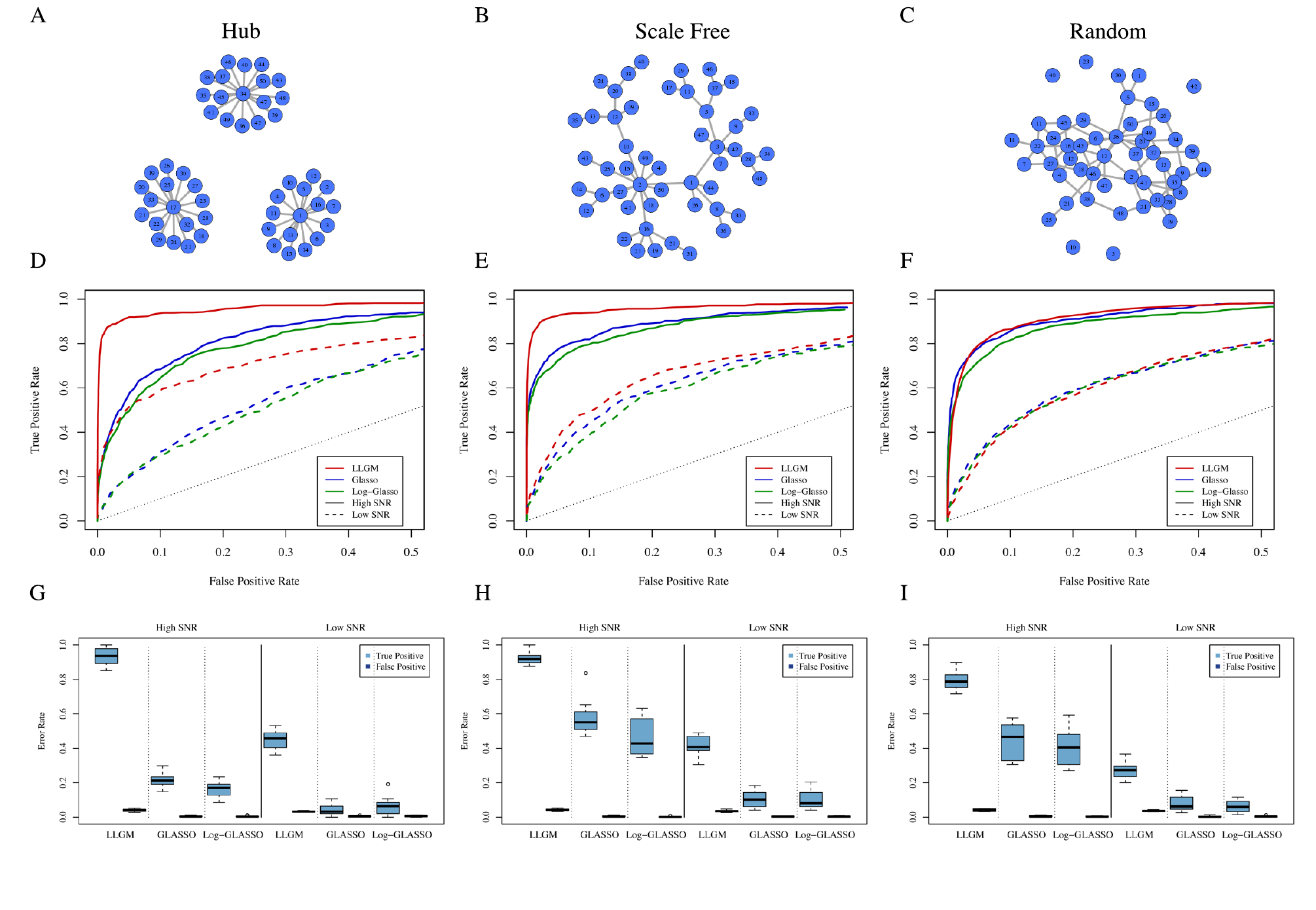}
\caption{ Experimental simulation study for three network structures:
  hub (A), 
  scale-free (B), and random (C).  For each type of graph, Poisson
  networks are generated with $200$ observations and $50$ nodes
  at a high and low signal-to-noise ratio (SNR).  Our {\em llgm} is
  compared to the 
  Graphical Lasso and the Graphical Lasso on the log-transformed data
  through Receiver Operator Curves (D-F) obtained by varying the
  regularization parameter, $\rho$, and boxplots (G - I) of true and
  false positive  rates (G-I) for fixed $\rho$ estimated by stability
  selection.   Our {\em llgm} outperforms competing methods for all three
  simulated network structures.      
 }\label{fig_sim}
\end{figure*}

We assess the performance of our {\em llgm} for
selecting the correct underlying network based on simulated count
data.  Three graph structures are
simulated: (i) a hub network, where each node is connected to one
of three hub nodes, (ii) a scale-free network, in which the number
of nodes of a certain degree follow a power law, and (iii) a random
network, in which each edge has equal probability.  The hub and
scale-free networks are known to mimic the behavior of biological
networks.  Our {\em llgm} is compared to the Graphical Lasso algorithm
\citep{glasso} and the Graphical Lasso after applying a log transform
to the data plus one.  Unlike for Gaussian graphical models and Ising
models, simulating Poisson networks is not a trivial task;
we employ an approach based on \citet{karlis_em_2003}.  In brief, $n$
independent observations from our simulated Poisson network
with $p$ nodes, $\X \in \Re^{n \times p}$, are
generated from the following model: $\X = \Y \mathbf{B} +
\mathbf{E}$.  Here, $\Y$ is a $n \times p + p(p-1)/2$ matrix with each
element $Y_{ij} \overset{iid}{\sim} Poisson( \lambda_{true})$ and
$\mathbf{E}$ is $n \times p$ with $E_{ij} \overset{iid}{\sim}
Poisson( \lambda_{noise})$.  The
matrix $\mathbf{B}$ encodes the true underlying graph structure
denoted by the adjacency matrix $\A \in \{ 0, 1 \}^{p \times p}$:
$\mathbf{B} = [ \mathbf{I}_{(p)} ; \mathbf{P} \odot ( \mathbf{1}_{(p)}
  \mathrm{tri}( \A)^{T} ) ]^{T}$.  Here, $\mathbf{P}$ is the $p
\times p(p-1)/2$ pair-wise permutation matrix, $\odot$ denotes the
Hadamard or element-wise product, and $\mathrm{tri}(\A)$ denotes the
$p(p-1)/2 \times 1$ vectorized upper triangular portion of the
adjacency matrix $\A$.  We simulate $n = 200$ observations for $p =
50$ nodes at two signal-to-noise (SNR) levels.  We set $\lambda_{true}
= 1$ with $\lambda_{noise} = 0.5$ for the high SNR level and
$\lambda_{noise} = 5$ for the low SNR level.

Results of our experiments conducted over ten replicates are
given in Figure \ref{fig_sim}. 
Both receiver operator curves (ROC) computed by varying the
regularization parameter $\rho$ and boxplots of true and
false positive rates for fixed $\rho$ as estimated via stability
selection are given at the high and low SNR levels.   True positives
are estimated as the fraction of edges found by {\em llgm} that are in the
true simulated network structure $\A$; false positives are estimated
analogously.  These results indicate that {\em llgm} uniformly outperforms
the Gaussian graphical models for the hub and scale-free graphs.  The
improved statistical power of our {\em llgm} for recovering the hub graph
structure is particularly striking.  The ROC curves of all methods on
the random graph structure are approximately equal.  When stability
selection is used to estimate the sparsity level, however, we
see that {\em llgm} retains its advantage over the competitors.  This behavior
is not surprising as employing the correct statistical model, in this
case the {\em llgm}, often leads to improved model selection.  Overall,
these simulation results demonstrate the strong performance of our
{\em llgm} for recovering network structures based on Poisson distributed
data.

\subsection{Discovering microRNA Networks}
\label{section_miRNA}

\begin{figure*}[!t]
\includegraphics[width=4.5in,clip=true,trim=.25in 0in 0in 0in]{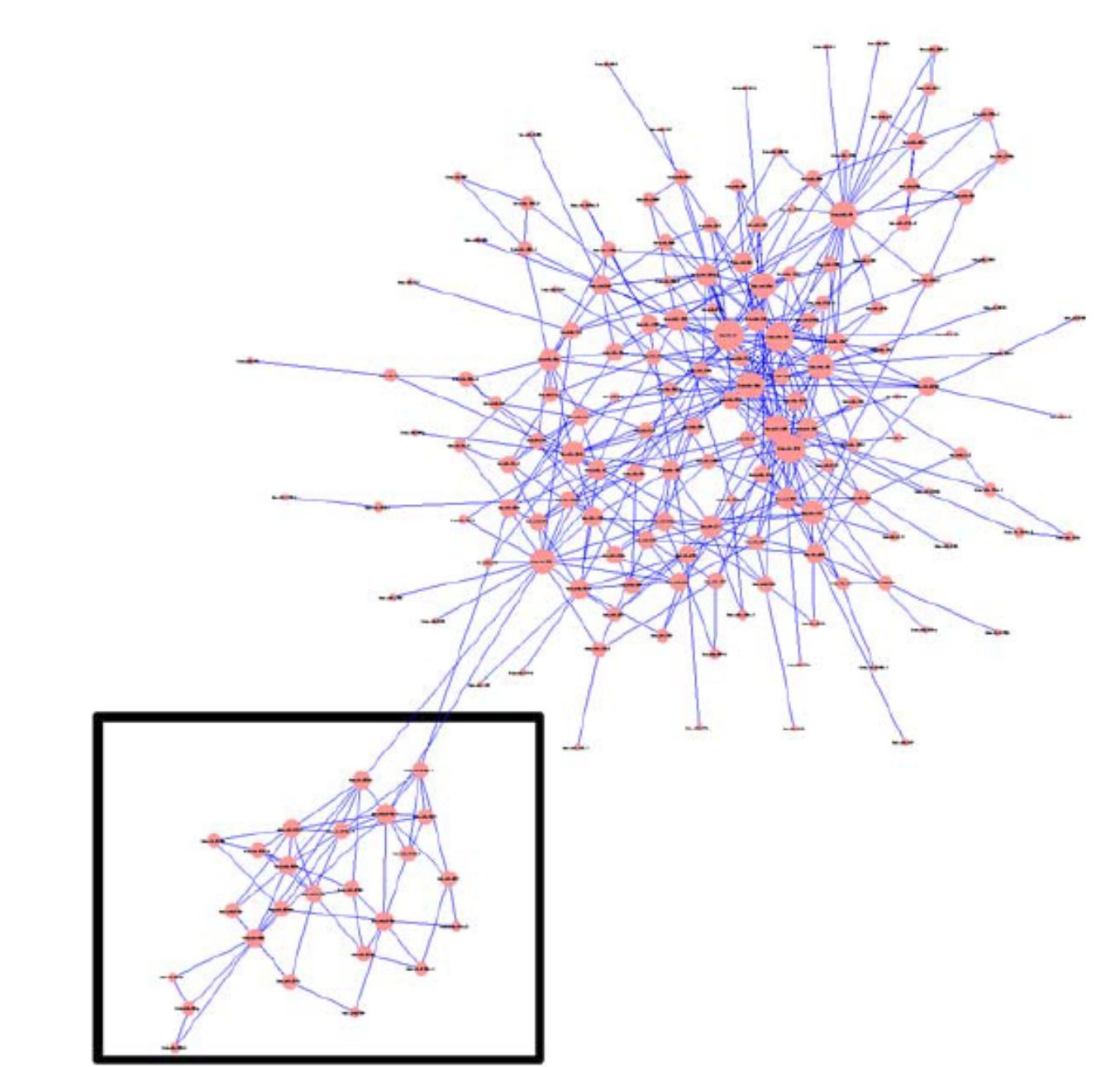}\includegraphics[width=2.5in]{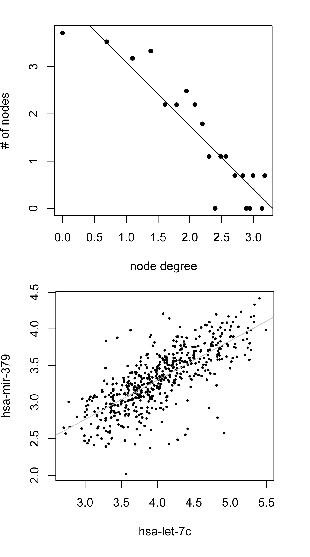} 
\caption{Breast cancer microRNA network estimated by {\em llgm} (left).  This
  network is scale-free as demonstrated by the power-law plot (top right) of
  node degree on the log-scale verses the number of nodes for such
  degree.  Our {\em llgm} found many hub genes previously associated
  with breast cancer such as let-7c, and identified new potential
  regulators of breast cancer such as mir-379 which is tightly correlated
  with let-7c (bottom right).  A microRNA cluster (left, boxed) was also
  identified by our 
  {\em llgm} in an unsupervised manner without using transcript location. 
}\label{fig_miRNA} 
\end{figure*}

We apply our {\em llgm} to discover relationships among microRNAs based on
sequencing data from breast cancer patients.  There is a long
record of applying Markov Networks to understand gene expression data,
but inferring networks based on microRNAs is a novel application of
graphical models.  Level III breast cancer
data was obtained from the Cancer Genome Atlas (TCGA) data portal
(http://tcga-data.nci.nih.gov/tcga/) \citep{collins_tcga_2007}.  This
data set consists of 544 patients and 524 microRNAs.  The sequencing
data was normalized as described in Section 2.1 with 50\% of the
microRNAs that varied the least across the samples filtered out,
giving us 262 microRNA nodes.

In Figure \ref{fig_miRNA}, we present the results of our {\em llgm} applied to
the breast cancer microRNA sequencing data.  Analysis of this network
reveals results consistent with the breast cancer genomics literature
as well as novel biomarkers and clusters to investigate further.
First, however, notice from the top right panel of Figure
\ref{fig_miRNA} that the estimated 
network closely follows a power law in the number of nodes at each
node degree, and thus appears to be a scale-free network.    Many
biological networks, such as gene expression networks, protein-protein
interaction networks, and metabolic networks, have been observed to be
scale-free 
\citep{barabsi_1999}; thus, we can add microRNA expression networks to
this list.

Many of the hub microRNAs identified in our
{\em llgm} such as let-7c, mir-10b, and mir-375 have been previously
associated with breast
cancer progression and metastasis.  For 
example,  let-7c has been shown to regulate the breast cancer metastatic
 \citep{Yu:2007bm}.   High level expression of
 mir-10b has been observed in triple negative, ER negative, PR
 negative, Her2 negative, breast 
cancer patients \citep{Radojicic:2011uf}.
Silencing mir-10b has been proposed as  
potential therapeutic target and tested in mouse mammary gland tumor models
\citep{Ma:2010cu}.  Blocking mir-375 in ER-positive 
cancer cells can slow down the cancer cell growth
\citep{deSouzaRochaSimonini:2010ev}.  Other hub microRNAs identified
in our {\em llgm} are novel biomarkers that need to be validated further for
associations in breast cancer.  Consider, for example,  mir-379 which
forms an edge and is tightly correlated, bottom right panel of Figure
\ref{fig_miRNA}, 
with another hub microRNA, let-7c.  There have been no studies on the
functionality of mir-379, but based on its hub status in the {\em llgm} and its
connections with other studied microRNAs, we hypothesize that mir-379
is a regulatory microRNA for breast cancer progression and
metastasis.

Our results also indicate interesting sub-network modules related to
microRNA clusters and functional 
regulatory pathways.  We identified a large microRNA cluster in the
right, boxed portion of Figure
\ref{fig_miRNA} which contains 
 has-mir-516a-1, 
has-mir-521-1, hsa-mir-522, has-mir-519a-1, and has-mir-527  
from chromosome 19:54251890-54265684 [+].
Many have established that microRNAs appear in
clusters on a single polycistronic transcript \citep{bentwich_2005}.   The
expression levels of precursor microRNAs in the same cluster are
synchronized and coordinated by similar transcription factors.  
Mature microRNAs levels, however, are regulated independently.  As 
sequencing technologies measure mature microRNA levels and we
did not incorporate any outside information such as transcript location, we
would not necessarily expect to find these microRNA clusters.
Interestingly, our {\em llgm} identifies a major microRNA cluster, indicating
that perhaps these microRNAs are functionally related, regulating similar
biological processes in breast cancer.

Overall, the novel application of our local Log-Linear Graphical Model to
understand breast cancer microRNA networks has yielded results
consistent with the known literature and identified potential
biomarkers and pathways for future research.

\section{Discussion}
\label{section_dis}

We have developed a novel framework for estimating
high-dimensional graphical models with Poisson distributed data.
While we have defined the global Poisson Markov Network, we have
chosen to estimate network structure locally, thus permitting a
richer set of dependencies among variables.  This is achieved through
fitting $\ell_{1}$ penalized log-linear models to select the
neighborhood of each node.  
Through simulations and a microRNA case study, we have demonstrated
the effectiveness of our {\em llgm} for estimating network structure from
high-throughput data.  

Our work leads to many areas of new methodological work.  These
include further studying the global {\em LLGM} and establishing
theoretical properties such as consistent graph recovery and
estimation of model parameters.  Also, a joint density defined on all
the nodes that does not severely restrict conditional relationships as
in the {\em LLGM} would be an important contribution.  Extensions of
graphical models to encompass other distributions is another direction
for future research.

There are many potential applications of our
Poisson graphical model and algorithm.  We have presented a case study on
microRNA networks, but clearly {\em llgm} will be useful for constructing
gene expression networks from RNA-sequencing data as well.  A major
consideration when applying our model to sequencing data is proper
normalization to ensure that the samples are (approximately)
independently and identically Poisson distributed.  In particular, as
our model is parametric, it is sensitive to zero-inflation and
overdispersion, both of which commonly occur with RNA-sequencing
data.  Thus, it is our strong recommendation to follow the
normalization steps described in Section \ref{section_Pois}; that is,
to filter out 
non-variable genes and adjust for
overdispersion. 
Additionally, our novel application using undirected graphs to study
microRNA networks yields a new method to examine microRNAs in groups
instead of the more common approaches of studying the genomic targets
of a single microRNA.  Further work to biologically validate our
predicted microRNA biomarkers and clusters in breast cancer is
needed to gain a more complete picture of the regulatory process in
this disease. 
Beyond genomics, there are many potential
applications of {\em llgm} to multivariate Poisson distributed data
such as that from 
user-ratings, web site visits, advertising clicks, bibliometrics, and
social networks.

In conclusion, our work developing Log-Linear Graphical Models for
high-throughput sequencing data has many implications and has opened
new directions for research both in the area of
high-dimensional graphical models and in the application of these to
gene expression and microRNA expression networks.

\section*{Acknowledgments}

The authors thank Pradeep Ravikumar for thoughtful insights and
helpful discussion related to this work.  
This work is supported in part through the  Collaborative Advances in
Biomedical Computing seed funding program at the Ken Kennedy Institute
for Information Technology 
at Rice University supported by the John and Ann Doerr Fund for
Computational Biomedicine and through the Center for Computational and
Integrative Biomedical Research Seed Funding Program at Baylor College
of Medicine.

\singlespacing

{\small

\bibliographystyle{Chicago}
\bibliography{network}

\begin{thebibliography}{}

\bibitem[\protect\citeauthoryear{Anders and Huber}{Anders and
  Huber}{2010}]{anders_2010}
Anders, S. and W.~Huber (2010).
\newblock Differential expression analysis for sequence count data.
\newblock {\em Genome Biol\/}~{\em 11\/}(10), R106.

\bibitem[\protect\citeauthoryear{Barab{\'a}si and Albert}{Barab{\'a}si and
  Albert}{1999}]{barabsi_1999}
Barab{\'a}si, A. and R.~Albert (1999).
\newblock Emergence of scaling in random networks.
\newblock {\em science\/}~{\em 286\/}(5439), 509.

\bibitem[\protect\citeauthoryear{Bentwich, Avniel, Karov, Aharonov, Gilad,
  Barad, Barzilai, Einat, Einav, Meiri, et~al.}{Bentwich
  et~al.}{2005}]{bentwich_2005}
Bentwich, I., A.~Avniel, Y.~Karov, R.~Aharonov, S.~Gilad, O.~Barad,
  A.~Barzilai, P.~Einat, U.~Einav, E.~Meiri, et~al. (2005).
\newblock Identification of hundreds of conserved and nonconserved human
  micrornas.
\newblock {\em Nature genetics\/}~{\em 37\/}(7), 766--770.

\bibitem[\protect\citeauthoryear{Besag}{Besag}{1974}]{besag_1974}
Besag, J. (1974).
\newblock Spatial interaction and the statistical analysis of lattice systems.
\newblock {\em Journal of the Royal Statistical Society. Series B
  (Methodological)\/}~{\em 36\/}(2), 192--236.

\bibitem[\protect\citeauthoryear{Bishop, Fienberg, and Holland}{Bishop
  et~al.}{2007}]{bishop_discrete_2007}
Bishop, Y., S.~Fienberg, and P.~Holland (2007).
\newblock {\em Discrete multivariate analysis}.
\newblock Springer Verlag.

\bibitem[\protect\citeauthoryear{Bullard, Purdom, Hansen, and Dudoit}{Bullard
  et~al.}{2010}]{bullard_rna_2010}
Bullard, J., E.~Purdom, K.~Hansen, and S.~Dudoit (2010).
\newblock Evaluation of statistical methods for normalization and differential
  expression in mrna-seq experiments.
\newblock {\em BMC bioinformatics\/}~{\em 11\/}(1), 94.

\bibitem[\protect\citeauthoryear{Collins and Barker}{Collins and
  Barker}{2007}]{collins_tcga_2007}
Collins, F. and A.~Barker (2007).
\newblock Mapping the cancer genome.
\newblock {\em Scientific American Magazine\/}~{\em 296\/}(3), 50--57.

\bibitem[\protect\citeauthoryear{de~Souza Rocha~Simonini, Breiling, Gupta,
  Malekpour, Youns, Omranipour, Malekpour, Volinia, Croce, Najmabadi,
  Diederichs, Sahin, Mayer, Lyko, Hoheisel, and Riazalhosseini}{de~Souza
  Rocha~Simonini et~al.}{2010}]{deSouzaRochaSimonini:2010ev}
de~Souza Rocha~Simonini, P., A.~Breiling, N.~Gupta, M.~Malekpour, M.~Youns,
  R.~Omranipour, F.~Malekpour, S.~Volinia, C.~M. Croce, H.~Najmabadi,
  S.~Diederichs, O.~Sahin, D.~Mayer, F.~Lyko, J.~D. Hoheisel, and
  Y.~Riazalhosseini (2010, November).
\newblock {Epigenetically deregulated microRNA-375 is involved in a positive
  feedback loop with estrogen receptor alpha in breast cancer cells.}
\newblock {\em Cancer research\/}~{\em 70\/}(22), 9175--9184.

\bibitem[\protect\citeauthoryear{Dempster}{Dempster}{1972}]{dempster_1972}
Dempster, A.~P. (1972).
\newblock Covariance selection.
\newblock {\em Biometrics\/}~{\em 28\/}(1), 157--175.

\bibitem[\protect\citeauthoryear{Dobra, Hans, Jones, Nevins, Yao, and
  West}{Dobra et~al.}{2004}]{dobra_2004}
Dobra, A., C.~Hans, B.~Jones, J.~Nevins, G.~Yao, and M.~West (2004).
\newblock Sparse graphical models for exploring gene expression data.
\newblock {\em Journal of Multivariate Analysis\/}~{\em 90\/}(1), 196--212.

\bibitem[\protect\citeauthoryear{Friedman, Hastie, and Tibshirani}{Friedman
  et~al.}{2007}]{glasso}
Friedman, J., T.~Hastie, and R.~Tibshirani (2007).
\newblock Sparse inverse covariance estimation with the lasso.
\newblock {\em Biostatistics\/}~{\em 9\/}(3), 432--441.

\bibitem[\protect\citeauthoryear{Friedman, Hastie, and Tibshirani}{Friedman
  et~al.}{2010}]{glmnet_paper_2010}
Friedman, J., T.~Hastie, and R.~Tibshirani (2010).
\newblock Regularization paths for generalized linear models via coordinate
  descent.
\newblock {\em Journal of Statistical Software\/}~{\em 33\/}(1), 1.

\bibitem[\protect\citeauthoryear{Hammersley and Clifford}{Hammersley and
  Clifford}{1971}]{hammersley_1971}
Hammersley, J. and P.~Clifford (1971).
\newblock Markov fields on finite graphs and lattices.

\bibitem[\protect\citeauthoryear{Hastie, Tibshirani, and Friedman}{Hastie
  et~al.}{2009}]{esl_2nd}
Hastie, T., R.~Tibshirani, and J.~J.~H. Friedman (2009).
\newblock {\em The elements of statistical learning\/} (2 ed.).
\newblock Springer.

\bibitem[\protect\citeauthoryear{Jiang and Wong}{Jiang and
  Wong}{2009}]{jiang_rna_2009}
Jiang, H. and W.~Wong (2009).
\newblock Statistical inferences for isoform expression in rna-seq.
\newblock {\em Bioinformatics\/}~{\em 25\/}(8), 1026--1032.

\bibitem[\protect\citeauthoryear{Karlis}{Karlis}{2003}]{karlis_em_2003}
Karlis, D. (2003).
\newblock An em algorithm for multivariate poisson distribution and related
  models.
\newblock {\em Journal of Applied Statistics\/}~{\em 30\/}(1), 63--77.

\bibitem[\protect\citeauthoryear{Kr{\"a}mer, Sch{\"a}fer, and
  Boulesteix}{Kr{\"a}mer et~al.}{2009}]{krmer_2009}
Kr{\"a}mer, N., J.~Sch{\"a}fer, and A.~Boulesteix (2009).
\newblock Regularized estimation of large-scale gene association networks using
  graphical gaussian models.
\newblock {\em BMC bioinformatics\/}~{\em 10\/}(1), 384.

\bibitem[\protect\citeauthoryear{Lauritzen}{Lauritzen}{1996}]{lauritzen_1996}
Lauritzen, S. (1996).
\newblock {\em Graphical models}, Volume~17.
\newblock Oxford University Press, USA.

\bibitem[\protect\citeauthoryear{Li, Witten, Johnstone, and Tibshirani}{Li
  et~al.}{2011}]{li_rna_2011}
Li, J., D.~Witten, I.~Johnstone, and R.~Tibshirani (2011).
\newblock Normalization, testing, and false discovery rate estimation for
  rna-sequencing data.
\newblock {\em Biostatistics\/}.

\bibitem[\protect\citeauthoryear{Liu, Roeder, and Wasserman}{Liu
  et~al.}{2010}]{liu_STARS_2010}
Liu, H., K.~Roeder, and L.~Wasserman (2010).
\newblock Stability approach to regularization selection (stars) for high
  dimensional graphical models.
\newblock {\em Arxiv preprint arXiv:1006.3316\/}.

\bibitem[\protect\citeauthoryear{Ma, Reinhardt, Pan, Soutschek, Bhat,
  Marcusson, Teruya-Feldstein, Bell, and Weinberg}{Ma et~al.}{2010}]{Ma:2010cu}
Ma, L., F.~Reinhardt, E.~Pan, J.~Soutschek, B.~Bhat, E.~G. Marcusson,
  J.~Teruya-Feldstein, G.~W. Bell, and R.~A. Weinberg (2010, April).
\newblock {Therapeutic silencing of miR-10b inhibits metastasis in a mouse
  mammary tumor model.}
\newblock {\em Nature biotechnology\/}~{\em 28\/}(4), 341--347.

\bibitem[\protect\citeauthoryear{Madigan, York, and Allard}{Madigan
  et~al.}{1995}]{madigan_1995}
Madigan, D., J.~York, and D.~Allard (1995).
\newblock Bayesian graphical models for discrete data.
\newblock {\em International Statistical Review/Revue Internationale de
  Statistique\/}~{\em 63\/}(2), 215--232.

\bibitem[\protect\citeauthoryear{Marioni, Mason, Mane, Stephens, and
  Gilad}{Marioni et~al.}{2008}]{marioni_2008}
Marioni, J., C.~Mason, S.~Mane, M.~Stephens, and Y.~Gilad (2008).
\newblock Rna-seq: an assessment of technical reproducibility and comparison
  with gene expression arrays.
\newblock {\em Genome research\/}~{\em 18\/}(9), 1509--1517.

\bibitem[\protect\citeauthoryear{Meinshausen and Buhlmann}{Meinshausen and
  Buhlmann}{2006}]{meinshausen_graphs_2006}
Meinshausen, N. and P.~Buhlmann (2006).
\newblock High-dimensional graphs and variable selection with the lasso.
\newblock {\em The Annals of Statistics\/}~{\em 34\/}(3), 1436--1462.

\bibitem[\protect\citeauthoryear{Mortazavi, Williams, McCue, Schaeffer, and
  Wold}{Mortazavi et~al.}{2008}]{mortazavi_2008}
Mortazavi, A., B.~Williams, K.~McCue, L.~Schaeffer, and B.~Wold (2008).
\newblock Mapping and quantifying mammalian transcriptomes by rna-seq.
\newblock {\em Nature methods\/}~{\em 5\/}(7), 621--628.

\bibitem[\protect\citeauthoryear{Radojicic, Zaravinos, Vrekoussis, Kafousi,
  Spandidos, and Stathopoulos}{Radojicic et~al.}{2011}]{Radojicic:2011uf}
Radojicic, J., A.~Zaravinos, T.~Vrekoussis, M.~Kafousi, D.~A. Spandidos, and
  E.~N. Stathopoulos (2011, February).
\newblock {MicroRNA expression analysis in triple-negative (ER, PR and
  Her2/neu) breast cancer.}
\newblock {\em Cell cycle (Georgetown, Tex.)\/}~{\em 10\/}(3), 507--517.

\bibitem[\protect\citeauthoryear{Ravikumar, Wainwright, and Lafferty}{Ravikumar
  et~al.}{2010}]{ravikumar_ising_2010}
Ravikumar, P., M.~Wainwright, and J.~Lafferty (2010).
\newblock High-dimensional ising model selection using l1-regularized logistic
  regression.
\newblock {\em The Annals of Statistics\/}~{\em 38\/}(3), 1287--1319.

\bibitem[\protect\citeauthoryear{Robinson and Oshlack}{Robinson and
  Oshlack}{2010}]{robinson_2010}
Robinson, M. and A.~Oshlack (2010).
\newblock A scaling normalization method for differential expression analysis
  of rna-seq data.
\newblock {\em Genome Biol\/}~{\em 11\/}(3), R25.

\bibitem[\protect\citeauthoryear{Wainwright and Jordan}{Wainwright and
  Jordan}{2008}]{wainwright_jordan_2008}
Wainwright, M. and M.~Jordan (2008).
\newblock Graphical models, exponential families, and variational inference.
\newblock {\em Foundations and Trends{\textregistered} in Machine
  Learning\/}~{\em 1\/}(1-2), 1--305.

\bibitem[\protect\citeauthoryear{Yu, Yao, Zhu, Zhang, Pan, Gong, Huang, Hu, Su,
  Lieberman, and Song}{Yu et~al.}{2007}]{Yu:2007bm}
Yu, F., H.~Yao, P.~Zhu, X.~Zhang, Q.~Pan, C.~Gong, Y.~Huang, X.~Hu, F.~Su,
  J.~Lieberman, and E.~Song (2007, December).
\newblock {let-7 regulates self renewal and tumorigenicity of breast cancer
  cells.}
\newblock {\em Cell\/}~{\em 131\/}(6), 1109--1123.

\bibitem[\protect\citeauthoryear{Yuan and Lin}{Yuan and
  Lin}{2007}]{yuan_lin_2007}
Yuan, M. and Y.~Lin (2007).
\newblock Model selection and estimation in the gaussian graphical model.
\newblock {\em Biometrika\/}~{\em 94\/}(1), 19.

\end{thebibliography}
}

\end{document}